\documentclass[twocolumn,showpacs]{revtex4}
\usepackage{graphicx}%
\usepackage{dcolumn}
\usepackage{amsmath}

\makeatletter
\def\btt#1{\texttt{\@backslashchar#1}}%
\DeclareRobustCommand\bblash{\btt{\@backslashchar}}%
\makeatother

\newcommand{\YPr}{Y$_{1-x}$Pr$_{x}$Ba$_2$Cu$_3$O$_{7-\delta}$~}
\newcommand{\Y}{YBa$_2$Cu$_3$O$_{7-\delta}$ }
\newcommand{\sixO}{$^{16}$O}
\newcommand{\eightO}{$^{18}$O}
\newcommand{\mSR}{$\mu$SR }
\newcommand{\Tc}{$T_{c}$ }

\newcommand{\Cel}{$^{o}$C }

\newcommand{\mab}{$m_{ab}^*$ }
\begin{document}
\preprint{PREPRINT (\today)}
\draft
%
%
\title{Oxygen-isotope effect on the
in-plane penetration depth in underdoped \YPr as revealed by
muon-spin rotation}
\author{R.~Khasanov$^{(1,2)}$, A.~Shengelaya$^{(1)}$, K.~Conder$^{(3)}$,
E.~Morenzoni$^{(2)}$, I.~M.~Savi\'c$^{(4)}$, and
H.~Keller$^{(1)}$}

\address {$^{(1)}$ Physik-Institut der Universit\"at Z\"urich, CH-8057
Z\"urich, Switzerland\\
$^{(2)}$ Paul Scherrer Institut, CH-5232 Villigen PSI, Switzerland\\
$^{(3)}$ Laboratory for Neutron Scattering, ETH Z\"urich and PSI
Villigen, CH-5232
Villigen PSI, Switzerland\\
$^{4}$ Faculty of Physics, University of Belgrade, 11001 Belgrade,
Yugoslavia}
%
%
%
%
\begin{abstract}
%
%
The oxygen-isotope (\sixO/\eightO) effect (OIE) on the in-plane
penetration depth $\lambda_{ab} (0)$ in underdoped \YPr was
studied by  muon-spin rotation. A pronounced OIE on
$\lambda_{ab}^{-2}(0)$ was observed with a relative isotope shift
of $\Delta\lambda^{-2}_{ab}/\lambda^{-2}_{ab}= -5(2)\%$ for  $x =
0.3$ and -9(2)\% for $x = 0.4$. It arises mainly from the
oxygen-mass dependence of the in-plane effective mass
$m_{ab}^{\ast}$. The OIE exponents of $T_{c}$ and of
$\lambda_{ab}^{-2}(0)$ exhibit a relation that appears to be
generic for cuprate superconductors.
\end{abstract}
~\\
\maketitle
%
%
\newpage
The pairing mechanism responsible for high-temperature
superconductivity remains elusive in spite of the  fact  that many
models have been proposed since its discovery. A fundamental
question is whether lattice effects are relevant for the
occurrence of high-temperature superconductivity. In order to
clarify this point a large number of isotope-effect studies were
performed since 1987 \cite{Franck94}. The first oxygen-isotope
effect (OIE) studies on the transition temperature $T_{c}$ were
performed on optimally doped samples, showing no significant
isotope shift \cite{Batlogg87}. However, later experiments
revealed a small but finite dependence of $T_{c}$ on the
oxygen-isotope mass $M_{\rm O}$
\cite{Batlogg87a,Cardona88,Crawford90,Zech94}, as well as on the
copper-isotope mass $M_{\rm Cu}$ \cite{Franck93,Zhao96}. Moreover,
a general trend in the dependence of the OIE exponent $\alpha_{\rm
O} = - d\ln T_c/d\ln M_{\rm O}$ on the doping level was found
which appears to be generic for all cuprate superconductors
\cite{Franck94,Crawford90,Franck91,Zhao96,Zhao98}: In the
underdoped region $\alpha_{\rm O}$ is large, even exceeding the
conventional BCS-value $\alpha = 0.5$ and becomes small in the
optimally doped and overdoped regime.

There is increasing evidence that a strong electron-phonon
coupling is present in cuprate superconductors, which may lead to
the formation of polarons (bare charge carriers accompanied by
local lattice distortions) \cite{Alexandrov94,Mueller2000}. One
way to test this hypothesis is to demonstrate that the effective
mass of the supercarriers $m^{*}$ depends on the mass $M$ of the
lattice atoms. This is in contrast to conventional BCS
superconductors, where only the `bare' charge carriers condense
into Cooper pairs, and $m^{*}$ is essentially independent of $M$.
For cuprate superconductors (clean limit) the in-plane penetration
depth $\lambda_{ab}$ is simply given by $\lambda_{ab}^{-2}(0)
\propto n_{s}/m_{ab}^{\ast}$, where $n_{s}$ is the superconducting
charge carrier density, and $m_{ab}^{\ast}$ is the in-plane
effective mass of the superconducting charge carriers. This
implies that the OIE on $\lambda_{ab}$ is due to a shift in
$n_{s}$ and/or $m_{ab}^{\ast}$:
\begin{equation}
\Delta\lambda^{-2}_{ab}(0)/\lambda^{-2}_{ab}(0)= \Delta n_s/n_s -\Delta
m_{ab}^{\ast}/m_{ab}^{\ast}.
\label{Deltalambda}
\end{equation}
Therefore a possible mass dependence of
$m_{ab}^{\ast}$ can be tested by investigating the isotope effect
on $\lambda_{ab}$, provided that the contribution of $n_{s}$ to the
total isotope shift is known.

Previous OIE studies of the penetration depth in \Y \cite{Zhao95},
La$_{2-x}$Sr$_{x}$CuO$_{4}$ \cite{Zhao97,Zhao98,Hofer2000}, and
Bi$_{1.6}$Pb$_{0.4}$Sr$_{2}$Ca$_{2}$Cu$_{3}$O$_{10+\delta}$~
\cite{Zhao2001} indeed showed a pronounced oxygen-mass dependence
on the supercarrier mass.  However, in all these experiments the
penetration depth was determined indirectly from the onset of
magnetization \cite{Zhao95,Zhao2001}, from the Meissner fraction
\cite{Zhao97,Zhao98}, and from magnetic torque measurements
\cite{Hofer2000}.  The muon-spin rotation (\mSR) technique is a
direct and accurate method to determine the penetration depth
$\lambda$ in type II superconductors.  In this Letter, we report
\mSR measurements of in-plane penetration depth $\lambda_{ab}$ in
underdoped \YPr ($x=0.3$ and 0.4) with two different oxygen
isotopes (\sixO~and \eightO). A large OIE on $\lambda_{ab}$ was
observed which mainly arises from the oxygen-mass dependence of
\mab.

Polycrystalline samples of \YPr ($x=0.3$ and $x=0.4$) were
prepared by standard solid state reaction
\cite{YPr123preparation}. Oxygen isotope exchange was performed
during heating the samples in $^{18}$O$_2$ gas. In order to ensure
the same thermal history of the substituted (\eightO) and not
substituted (\sixO) sample, two experiments (in $^{16}$O$_2$ and
$^{18}$O$_2$) were always performed simultaneously.  The exchange
and back exchange processes were carried out at 600\Cel during
25~h, and then the samples were slowly cooled (20$^o$C/h) in order
to oxidize them completely.  The \eightO~content in the samples,
as determined from a change of the sample weight after the isotope
exchange, was found to be 78(2)\% for both samples.  The total
oxygen content of the samples was determined using high-accuracy
volumetric analysis \cite{YPr123preparation}. To examine the
quality of the samples low-field (1mT, field-cooled) SQUID
magnetization measurements were performed (see Fig.~1). For both
concentrations the \Tc onset for the \sixO~samples was higher than
for \eightO~with nearly the same transition width.  An oxygen back
exchange of the \eightO~sample ($x = 0.4$) resulted within error
in almost the same magnetization curve as for the \sixO~ sample,
confirming that the back exchange is almost complete. The results
of the OIE on $T_{c}$ are summarized in Table I.  Taking into
account an isotope exchange of 78\%, we found $\alpha_{\rm O}$ =
0.22(4) for $x =0.3$ and $\alpha_{\rm O}$ = 0.37(5) for $x =0.4$,
in agreement with previous results \cite{Franck91,Zhao96a}.

\begin{figure}[htb]
\includegraphics[width=0.9\linewidth]{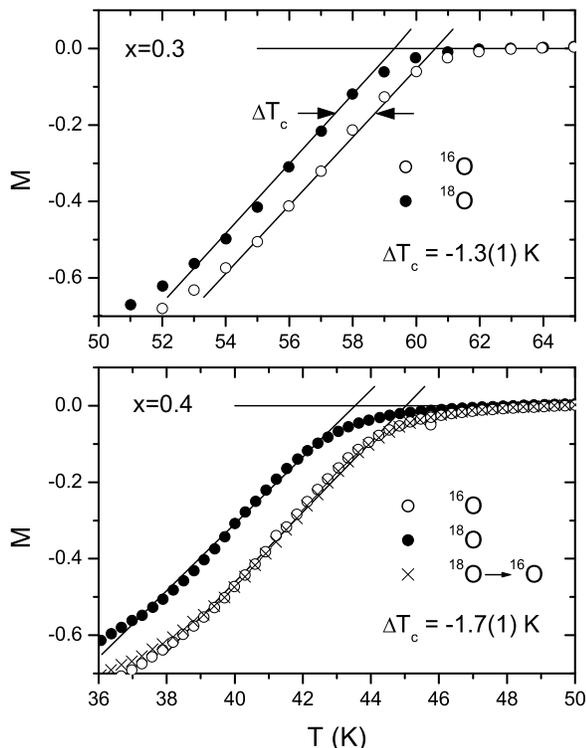}
\vspace {-0.5cm} \caption{Section near T$_c$ of the low-field
(1mT, field-cooled) magnetization curves (normalized to the value
at 10K) for  \YPr (x~=~0.3 and 0.4).}
\end{figure}

The \mSR experiments were performed at the Paul Scherrer Institute
(PSI), Switzerland, using the $\pi$M3 \mSR facility.  The samples
consisted of sintered pellets ( 12~mm in diameter, 3~mm thick)
which were mounted on a Fe$_{2}$O$_{3}$ sample holder in order to
reduce the background from muons not stopping in the sample.  The
polycrystalline \YPr samples were cooled from far above $T_{c}$ in
a magnetic field of 200~mT perpendicular to the sample disk.
Time-differential \mSR spectroscopy was employed, from which one
can deduce the probability distribution of the local magnetic
field $p(B)$ of the vortex state by measuring the time evolution
of the muon-spin polarization \cite{Lee99}.  In a powder sample
the magnetic penetration depth $\lambda$ can be extracted from the
muon-spin depolarization rate $\sigma(T) \propto
1/\lambda^{2}(T)$, which probes the second moment $\langle \Delta
B^{2}\rangle^{1/2}$ of $p(B)$ in the mixed state
\cite{Lee99,Zimmermann95}.  For highly anisotropic layered
superconductors (like the cuprate superconductors) $\lambda$ is
mainly determinated by the in-plane penetration depth
$\lambda_{ab}$ \cite{Zimmermann95}: $ \sigma(T) \propto
1/\lambda_{ab}^{2}(T) \propto n_{s}/m_{ab}^{\ast}.$

\begin{figure}[htb]
\includegraphics[width=0.9\linewidth]{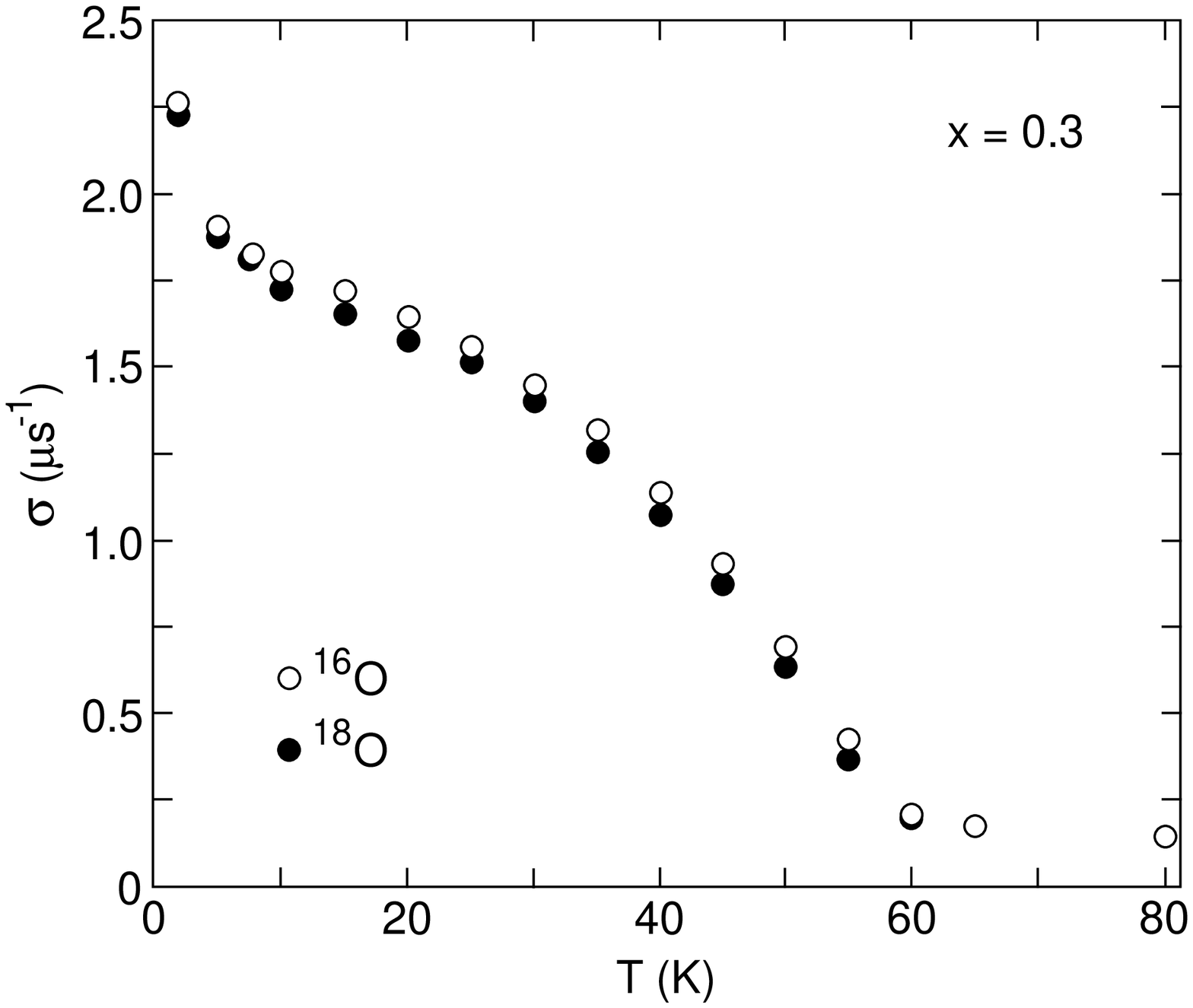}
\vspace{-0.4cm} \caption{Temperature dependence of  the $\mu$SR
depolarization rate $\sigma$ of \YPr for $x =0.3$, measured in a
field 200~mT (field-cooled). }
\end{figure}

The depolarization rate $\sigma$ was extracted from the \mSR time
spectra using a Gaussian relaxation function $R(t) =
\exp[-\sigma^{2}t^{2}/2]$. Figure~2 shows the temperature
dependence of the measured $\sigma$ for the \YPr samples with $x =
0.3$.  Similar results were obtained for the samples with $x =
0.4$. It is evident that the values of $\sigma$ for \eightO~ are
systematically lower than those for \sixO. As expected for a type
II superconductor in the mixed state, $\sigma$ continuously
increases below $T_{c}$ with decreasing temperature
\cite{Zimmermann95}. The sharp increase of $\sigma$ below $\simeq
10$~K is due to antiferromagnetic ordering of the Cu(2) moments
\cite{Seaman90}. Above $T_{c}$ a small temperature independent
depolarization rate $\sigma_{nm} \simeq 0.15 \; \mu {\rm s}^{-1}$
is seen, arising from the nuclear magnetic moments of Cu and Pr.
Therefore, the total $\sigma$ is determined by three
contributions: a superconducting ($\sigma_{sc}$), an
antiferromagnetic ($\sigma_{afm}$), and a small nuclear magnetic
dipole ($\sigma_{nm}$) contribution. Because $\sigma_{afm}$ is
only present at low temperatures, data points below 10~K were not
considered in the analysis. The superconducting contribution
$\sigma_{sc}$ was then determined by subtracting $\sigma_{nm}$
measured above $T_{c}$ from $\sigma$. In Fig.~3 we show the
temperature dependence of $\sigma_{sc}$ for the \YPr samples with
$x = 0.3$ and 0.4.  It is evident that for both concentrations a
remarkable oxygen isotope shift on \Tc as well as on $\sigma_{sc}$
is present.

\begin{figure}[htb]
\includegraphics[width=0.95\linewidth]{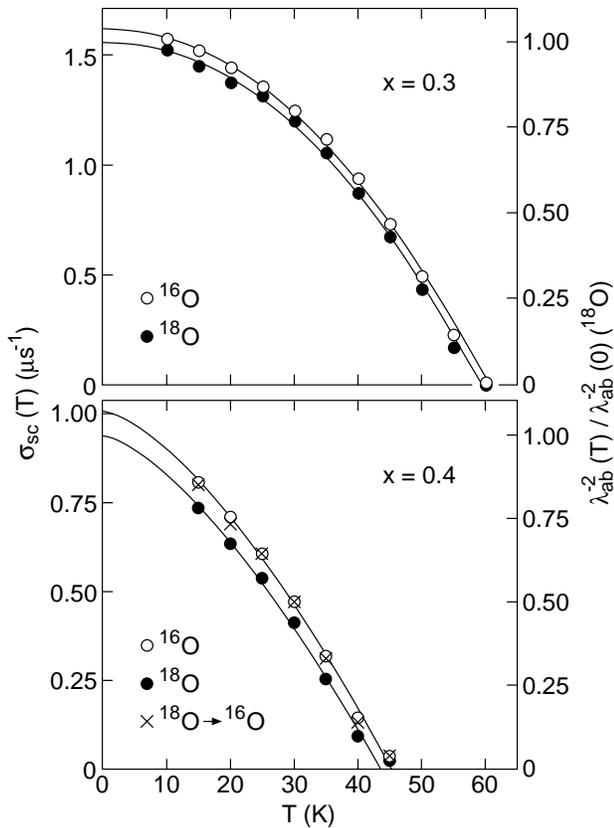}
\vspace{-0.4cm} \caption{Temperature dependence of depolarization
rate $\sigma_{sc}$ in \YPr  for $x = 0.3$ and $0.4$ (200~mT,
field-cooled).  On the right axis the normalized in-plane
penetration depth
$\lambda^{-2}_{ab}(T)/\lambda^{-2}_{ab}(0)$(\eightO) is plotted
for comparison with Ref.~\cite{Hofer2000}. The solid lines
correspond to fits to the power law
$\sigma_{sc}(T)/\sigma_{sc}(0)=1- (T/T_{c})^n$. }
\end{figure}
The data in Fig.~3 were fitted to the power law
$\sigma_{sc}(T)/\sigma_{sc}(0)=1- (T/T_{c})^n$ \cite{Zimmermann95}
with $\sigma_{sc}(0)$ and $n$ as free parameters, and \Tc fixed.
The values of \Tc were taken from the magnetization measurements
(see Table I).  The values of $\sigma_{sc}(0)$ obtained from the
fits are listed in Table I and are in agreement with previous
results \cite{Seaman90}.  The exponent $n$ was found to be $n =
2.0(1)$ for $x = 0.3$ and $n = 1.5(1)$ for $x = 0.4$, which is
typical for underdoped YBCO \cite{Zimmermann95}. Moreover, $n$ is
within error the same for \sixO~ and \eightO. This implies that
$\sigma_{sc}$ has nearly the same temperature dependence for the
two isotopes (see Fig.~3).  In order to proof that the observed
OIE on $\lambda_{ab}(0)$ are intrinsic, the \eightO~ sample with
$x = 0.4$ was back exchanged (\eightO~$\to$ \sixO). As seen in
Fig.~3, the data points of this sample (cross symbols) indeed
coincide with those of the \sixO~ sample. From the values of
$\sigma_{sc}(0)$ listed in Table I the relative isotope shift of
the in-plane penetration depth $\Delta
\lambda_{ab}^{-2}(0)/\lambda_{ab}^{-2}(0) = [\sigma_{sc}^{\rm
18O}(0) - \sigma_{sc}^{\rm 16O}(0)] /\sigma_{sc}^{\rm 16O}(0)$ was
determined. Taking into account an isotope exchange of 78\%, one
finds  $\Delta\lambda^{-2}_{ab}(0)/\lambda^{-2}_{ab} (0)= -5(2)\%$
and -9(2)\%  for $x=0.3$ and 0.4, respectively (Table I). For the
OIE exponent $\beta_{\rm O} = - d\ln \lambda_{ab}^{-2}(0)/d\ln
M_{\rm O}$, one readily obtains $\beta_{\rm O}$ = 0.38(12) for $x
=0.3$ and $\beta_{\rm O}$ = 0.71(14) for $x =0.4$ (Table I). This
means that in underdoped \YPr the OIE on $\lambda_{ab}^{-2}$ as
well as on \Tc increase with increasing Pr doping $x$ (decreasing
\Tc). This finding is in excellent agreement with the recent
magnetic torque measurements on underdoped
La$_{2-x}$Sr$_{x}$CuO$_{4}$ \cite{Hofer2000}.
\begin{table}
\caption[~]{Summary of the OIE results for \YPr extracted from the
experimental data (see text for an explanation).  } %
\begin{center}
\begin{tabular}{lllccccccc} \hline\hline
   &\multicolumn{2}{c}{\sixO} &\multicolumn{2}{c}{\eightO} &&
& && \\
\hline $x$ &\Tc& $\sigma_{sc}(0)$ &\Tc& $\sigma_{sc}(0)$
&$\alpha_{\rm O}$&
$\frac{\Delta \lambda_{ab}^{-2}(0)}{\lambda_{ab}^{-2}(0)}$ &$\beta_{\rm O}$ \\
 &[K]& [$\mu$s$^{-1}$] &[K]& [$\mu$s$^{-1}$] && [\%] &\\
\hline
0.3 &60.6(1)& 1.63(2) &59.3(1)& 1.57(2)  &0.22(4)& -5(2) &  0.38(12) \\
0.4 &45.3(1)& 1.01(2) &43.6(1)& 0.94(2)  &0.37(5)& -9(2) & 0.71(14) \\
0.4& 45.1(1)\footnotemark[1]& 1.01(4)\footnotemark[1]&&&&& \\
\hline
\hline
\end{tabular}
\footnotetext[1]{results for the back-exchange (\eightO $\to$
\sixO) sample }
\end{center}
\end{table}

According to Eq.~(\ref{Deltalambda}) the observed
$\Delta\lambda_{ab}^{-2}(0)/\lambda_{ab}^{-2}(0)$ is due to a
shift of $n_{s}$ and/or $m_{ab}^{\ast}$. For
La$_{2-x}$Sr$_{x}$CuO$_{4}$ several independent experiments
\cite{Zhao97,Zhao98,Hofer2000} have shown that the change of
$n_{s}$ during the exchange procedure is negligibly small. In the
present work we provide further evidence: (i) The fully oxygenated
\YPr samples ($\delta \simeq 0$) were all prepared under identical
conditions, either in a $^{16}{\rm O}_{2}$ or $^{18}{\rm O}_{2}$
atmosphere \cite{YPr123preparation}, and the Pr content $x$ did
not change. It is very unlikely that $n_{s}$ changes significantly
upon \eightO~ substitution, and after the back-exchange (\eightO
$\to$ \sixO) the same results are obtained (see Figs.~1, 3 and
Table I). (ii) According to a model \cite{Lichtenstein95} that
describes the suppression of \Tc in
Y$_{1-x}$Pr$_{x}$Ba$_2$Cu$_3$O$_{7-\delta}$, the number of
supercarriers decreases linearly with increasing $x$ in the range
of $0.05 < x < 0.5$, and consequently $\Delta n_{s}/n_{s}= -\Delta
x/x$. Moreover, for $0.1< x <0.5$ the transition temperature \Tc
decreases linearly with $x$, with $\Delta T_c/\Delta x \simeq
-150$~K/Pr atom \cite{Franck91}. Combining this two relations one
obtains: $\Delta T_{c} \simeq -150\cdot x \cdot\Delta
n_{s}/n_{s}$. Assuming that the observed OIE on
$\lambda^{-2}_{ab}$ is only due to a change of $n_{s}$ ($\Delta
m_{ab}^{\ast}/m_{ab}^{\ast} \simeq 0$), one can estimate the
corresponding shift of $T_{c}$. For $x = 0.3$ and $x = 0.4$ one
finds $\Delta T_{c} \simeq - 1.8(4)$~K and $- 4.2(6)$~K,
respectively. The experimental values, however, are much lower
(see Fig.~1): $\Delta T_{c} = - 1.3(1)$~K ($x = 0.3$) and $\Delta
T_{c} = - 1.7(1)$~K ($x = 0.4$).  We thus conclude that any change
in $n_{s}$ during the exchange procedure must be small, and that
the change of $\lambda_{ab}$ is mainly due to the OIE on the
in-plane effective mass $m_{ab}^{\ast}$ with $\Delta
m_{ab}^{\ast}/m_{ab}^{\ast} \simeq 5(2)$~\% and 9(2)~\% for $x =
0.3$ and $x = 0.4$, respectively.  This implies that the effective
supercarrier mass $m_{ab}^{\ast}$ in this cuprate superconductor
depends on the oxygen mass of the lattice atoms, which is not
expected for a conventional phonon-mediated BSC superconductor.

In Fig.~4 the exponent $\beta_{\rm O}$ versus the exponent
$\alpha_{\rm O}$ for \YPr~ is plotted. For comparison we also
included the recent magnetic torque results of underdoped
La$_{2-x}$Sr$_{x}$CuO$_{4}$ \cite{Hofer2000}.  It is evident that
these exponents are linearly correlated: $\beta_{\rm O} =
A\cdot\alpha_{\rm O} + B$. A best fit yields $A = 1.8(4)$ and $B =
-0.01(12)$, so that $\beta_{\rm O} \simeq A\cdot\alpha_{\rm O}$.
This empirical relation appears to be generic for cuprate
superconductors. Quantitatively one can understand this behavior
in terms of an empirical relation between \Tc and the \mSR
depolarization rate $\sigma_{sc}(0)$
\cite{Uemura89,SchneiderKeller92}. It was shown
\cite{SchneiderKeller92} that for most families of cuprate
superconductors the simple parabolic relation
$\overline{T}_{c}=2\overline{\sigma}(1-\overline{\sigma}/2)$
describes the experimental data rather well (here
$\overline{T}_{c}=T_{c}/T^{m}_{c}$,
$\overline{\sigma}=\sigma_{sc}(0)/\sigma_{sc}^{m}(0)$, and
$T_{c}^{m}$ and $\sigma_{sc}^{m}(0)$ are the transition
temperature and depolarization rate of the optimally doped
system). Using this parabolic Ansatz, one readily obtains the
linear relation between $\beta_{\rm O}$ and $\alpha_{\rm O}$:
$\beta_{\rm O} /\alpha_{\rm O} = 1 + 1/2 \; [(1-(1-
\overline{T}_{c})^{1/2})/(1-\overline{T}_{c})^{1/2}]$.
\begin{figure}[htb]
\includegraphics[width=0.8\linewidth]{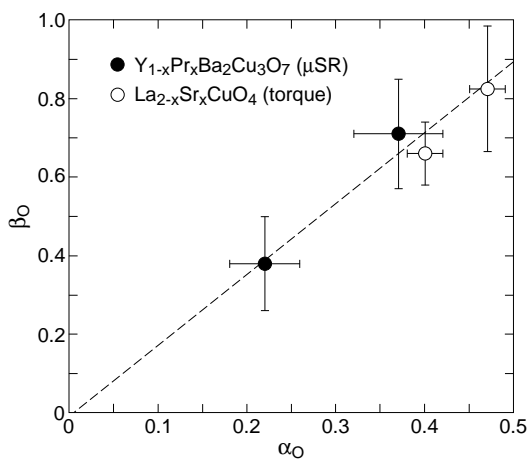}
\vspace{-0.4cm} \caption{Plot of the OIE exponents $\beta_{\rm O}$
versus $\alpha_{\rm O}$ for \YPr ($ x = 0.3$ and $0.4$) and
La$_{2-x}$Sr$_{x}$CuO$_{4}$ ($x = 0.080$ and $0.086$) from
\cite{Hofer2000}. The dashed line represents a best fit to the
empirical relation $\beta_{\rm O} = A\cdot\alpha_{\rm O} + B$.}
\end{figure}
In the heavily underdoped regime ($\overline{T}_{c} \rightarrow 0
$) $\beta_{\rm O} /\alpha_{\rm O}  \rightarrow 1$.  For the
underdoped samples shown in Fig.~4 the reduced critical
temperature $\overline{T}_{c}$ is in the range 0.5 to 0.7,
yielding $\beta_{\rm O} /\alpha_{\rm O} = 1.2 - 1.4$, in agreement
with $A = 1.8(4)$ obtained from the experimental data. Very
recently, it was pointed out \cite{SchneiderKeller2001} that the
unusual doping dependence of the OIE on \Tc and on
$\lambda_{ab}^{-2}(0)$ naturally follows from the doping driven
3D-2D crossover and the 2D quantum superconductor to insulator
transition in the underdoped limit. It is predicted that in the
underdoped regime $\beta_{\rm O} / \alpha_{\rm O} \rightarrow 1$,
which is consistent with the parabolic Ansatz.

In summary, we performed \mSR measurements of the in-plane
penetration depth $\lambda _{ab}$ in underdoped \YPr ($x = 0.3,
0.4$) for samples with two different oxygen isotopes ($^{16}$O and
$^{18}$O).  A pronounced OIE on both the transition temperature
\Tc and $\lambda_{ab}^{-2}(0)$ was observed, which increases with
decreasing $T_{c}$.  The isotope shift on $\lambda_{ab}^{-2}(0)$
is attributed to a shift in the in-plane effective mass
$m_{ab}^{\ast}$.  For $x = 0.3$ and 0.4 we find $\Delta
m_{ab}^{\ast}/m_{ab}^{\ast} = -5(2)\%$ and -9(2)\%, respectively.
Furthermore, an empirical relation between the OIE exponents
$\beta_{\rm O}$ and $\alpha_{\rm O}$ was found that appears to be
generic for various classes of cuprate superconductors.  The OIE
on $m_{ab}^{\ast}$ implies that the superconducting carriers have
polaronic character, and that lattice effects play an essential
role in the occurrence of high-temperature superconductivity.

We are grateful to G.M.~Zhao, T.~Schneider, and K.A.~M\"{u}ller
for many fruitful discussions and to A.~Amato, U.~Zimmermann, and
D.~Herlach from PSI for technical support during the \mSR
experiments.  This work was partly supported by the Swiss National
Science Foundation.


%
\end{document}